\documentclass[%
aip,
amsmath,amssymb,
reprint,%
]{revtex4-1}

\usepackage{graphicx}
\usepackage{dcolumn}
\usepackage{bm}

\usepackage[utf8]{inputenc}
\usepackage[T1]{fontenc}
\usepackage{mathptmx}
\usepackage{etoolbox}

\makeatletter
\def\@email#1#2{%
	\endgroup
	\patchcmd{\titleblock@produce}
	{\frontmatter@RRAPformat}
	{\frontmatter@RRAPformat{\produce@RRAP{*#1\href{mailto:#2}{#2}}}\frontmatter@RRAPformat}
	{}{}
}%
\makeatother
\begin{document}
	
	\preprint{AIP/123-QED}
	
	\title[]{Nonlinear dynamical social and political prediction algorithm for city planning and public participation using the Impulse Pattern Formulation}
	\author{R. Bader}
	\email{R\_Bader@t-online.de.}
\affiliation{Institute of Systematic Musicology, University of Hamburg, Neue Rabenstr. 13, 20354 Hamburg}
\author{S. Linke}%
\affiliation{ 
Hamburg University for Applied Sciences, Ligeti Center, Veritaskai 1, 21079 Hamburg
}%

\author{S. Gernert}
\affiliation{%
Freie und Hansestadt Hamburg
Landesbetrieb Immobilienmanagement und Grundvermögen
Millerntorplatz 1
20359 Hamburg \\
}%
	
\date{\today}

\begin{abstract}
	A nonlinear-dynamical algorithm for city planning is proposed as an Impulse Pattern Formulation (IPF) for predicting relevant parameters like health, artistic freedom, or financial developments of different social or political stakeholders over the cause of a planning process. The IPF has already shown high predictive precision at low computational cost in musical instrument simulations, brain dynamics, and human-human interactions. 
	The social and political IPF consists of three basic equations of system state developments, self-adaptation of stakeholders, two adaptive interactions, and external impact terms suitable for respective planning situations. Typical scenarios of stakeholder interactions and developments are modeled by adjusting a set of system parameters. These include stakeholder reaction to external input, enhanced system stability through self-adaptation, stakeholder convergence due to adaptive interaction, as well as complex dynamics in terms of fixed stakeholder impacts. A workflow for implementing the algorithm in real city planning scenarios is outlined. This workflow includes machine learning of a suitable set of parameters suggesting best-practice planning to aim at the desired development of the planning process and its output.
\end{abstract}

\maketitle

\begin{quotation}
	Modern city planning, as well as social or political decision-making, depend on predictions of the outcome of actions, which are mainly based on statistical data or expert opinions. Still, the complexity of modern societies is steadily increasing while resources are getting scarce. Also, public participation has become increasingly important, as modern city planning is often based on the upcycling of materials and increased densification of cities rather than the planning of new urban areas. To increase predictive robustness, efficient use of resources, and social participation and public acceptance of planned actions, a physical modeling algorithm is proposed. Such an algorithm needs to be computationally cheap to allow large-scale system prediction in a reasonable time. It needs to account for typical planning situations with sudden phase changes, adaptation, and mutual interactions, and it needs to be able to connect stakeholders in a planning process to be modeled with parameters relevant to these stakeholders like health, artistic freedom, social peace, or financial feasibility. The algorithm proposed is, therefore, based on a Physical Culture Theory, taking physics as a common ground. Such an algorithm also needs to be able to predict scenarios that have not been present before. Therefore, physical modeling is superior to a machine learning algorithm, which only knows what it has learned. Such an algorithm could, in the long run, be used in public participation processes on a web-based modification and suggestion tool that is freely available online.
\end{quotation}

\section{Introduction}

"We are already colourful" A statement that does not seem to fit into the scientific context and yet has led to the introduction of an algorithm in urban planning to test, question or confirm this statement, for a review see \cite{Abas2023}. The statement of the citizens who see themselves exposed to the plans for a future change in their neighbourhood. How can urban planners advance their projects in a targeted manner in the interests of the community and expect a "Yes please, let's examine the parameters for a liveable future together" instead of a "No thanks, we've already set up our lives"? Many approaches have been proposed mitigating this problem, often involving machine learning algorithms.

In times of scarcity of space and land, diverging political goals such as the net zero land strategy and alliances in favour of building even more housing, the smallest parameters that are not taken into account can have a decisive influence on the already fragile balance and even prevent planning altogether\cite{Charturvedi2021}. Active citizens' initiatives can lead to referendums and long-term uncertainties in urban planning. On both sides of the process, resources and lifetimes are spent fighting against their own system. There is no question that planning in an increasingly dense city is becoming more complex and planning for the future more uncertain. But building takes time and what is built is there for the time being: it creates space and thus shapes society. Both the new neighbourhood and the old neighbourhood need to be well planned or replanned, taking into account all known parameters, so that stable, sustainably communities and thus stable societies can form. With forward-looking planning that takes environmental concerns, demographic aspects and crises into account, an equal society is conceivable.\cite{Wilkinson2024}.

The world is more colourful than a single planning team can grasp. An algorithm can provide support here and visualise how complex living beings and things behave. However, it can show citizens just as credibly that this is a forward-looking and necessary decision by the planners. 

One example of a complex planning situation that involved and activated many stakeholders was the replanning of the "Frappant" shopping centre in Hamburg Altona, Germany, which was built in 1973. Since the 1980s, the huge "concrete block" has been in decline and stood completely empty for several years until the 2005. Then, one by one, artists moved in as temporary tenants and the city declared the neighbourhood a redevelopment area. By 2009, 150 artists had come together to form the "frappant.org" initiative and were using the spaces with different aims. Some appreciate the large and affordable studios, others want to establish a new cultural centre for the city. In autumn 2009, IKEA buys the complex to establish a city centre branch. Citizens' movements for (e.g. business men) and against (e.g. local residents) IKEA are formed. The old Frappant building is demolished at the end of 2010\cite{Necker2014}. 

Considering self-organizing the foundation of social and political systems and therefore claiming a need for a physics- and nature-based politics has prominently been advocated for on philosophical grounds\cite{Latour2022, Latour2017}. These ideas have been formulated in a much broader sense as ecosystems and life in general to be of self-organizing nature\cite{Margulis1998,Lovelock1988,Smith2016,Watson1983}.

Following this road, nonlinear dynamical models for political and social issues have extensively been proposed in terms of social complexity measures (see\cite{Flack2011} for a review), smart city infrastructure\cite{Rajaan2024}, power grids energy management stability analysis using machine learning\cite{Titz2024}, or low-carbon transport city planning\cite{Xinghui2012}, next to many others. The need to rethink city planning and urbanism is felt strongly in the fields of city planning and architecture in the direction of using sustainable materials and ecological urbanism\cite{Tucci2022,Coskun2023}.

Social polarization has often been studied using different nonlinear dynamical methods. Using a Hamiltonian approach of connected social network coherence minimization, a phase transition was found when the amount of connectivity in the society rises above a certain level, which only has happened through extensive internet connectivity over the last decades\cite{MinhPham2020}. This model allows only a polar decision regarding friend/enemy. The role of tipping points has been stressed in a model of hysteresis loop polaization processes\cite{Macya2022}. This model is agent-based rather than using differential equations. Another approach is using statistical methods like Boltzman-type dynamics\cite{Loy2022}. This allows for analytical descriptions of phase-changes in polarization processes.  

Using a time-step method of individual opinion development changing according to the impact of all other 
network participants, coherent and polarized social states can be found by alternating the impact strength and the shape of a hyperbolic tangent function\cite{Baumann2020}. This model enforces polarization due to positive or negative opinions from the start but allows the opinions to be continuously moderate to extreme.

In sociology, the structural balance theory initially assumed a set of three members being friends or enemies and changing relations according to a balanced or unbalanced state\cite{Heider1946}. It has been enlarged to large societies\cite{Taylor1970}. In a model of a set of differential equations for homogeneous or diverging societies, it was found that initial model conditions determined the output, where initial mean positive relations lead to conformity. In contrast, initial mean hostile relations lead to divergence\cite{Marvel2011}.

Conformity simulation in consumer models shows a relation between conformity network members' stability depending on the network topology\cite{Jedrzejewski2020}. Conformity stability is much more stable with balanced local/global connections than with an unbalanced network.

All these models are not perfectly suitable for the task of city planning. There, different stakeholders participating in a planning process have different interests like social balance, artistic freedom, financial interest, etc. These have an initial state at the start and will change throughout the construction process due to complex interactions among stakeholders, internal adaptation processes, or external influences. Therefore, a dynamic system must be formulated based on all stakeholders and their interactions.

Public participation is crucial in deciding if and how a planning process needs to take place. During this stage, different opinions exist on the impact of political action on different stakeholders. Some might find the social or financial balance in a city environment at risk, shop owners might fear financial downgrading, and artists might fear losing their performance or exhibition spaces.

Therefore, a robust and persuasive prediction method is needed to estimate the influence of political and social measures on all stakeholders. Public participation could agree on the desired output of the measures. According to this aim, a nonlinear dynamical network is needed to predict the best measures to be taken to achieve this aim.

The Impulse Pattern Formulation (IPF) model proposed in the present paper to arrive at this aim has first been developed for musical instrument applications\cite{Bader2013,Linke2021b,Linke2019b}. It has also been proven to predict human-human interaction in rhythm perception and production\cite{Linke2021a}, according to previous physiological\cite{Buzsaki2006}\cite{Ha2017} and model findings\cite{Fuhrmann2002}\cite{Fardet2020}. The IPF has also proven to work in brain dynamics\cite{Bader2022}, explaining fundamental aspects like the excitatory/inhibitory neuron concentration or typical brain time scales. The brain IPF also corresponds to EEG measurements of brain synchronization as the physiological cause of musical large-scale tension dynamics\cite{Bader2024}.

The IPF takes a viewpoint neuron of an object from which an impulse is sent out to several other neurons, musical instrument parts, or any object. This impulse is processed, damped, and returns to the viewpoint object\cite{Bader2013,Linke2019b,Bader2021,Linke2021b} leading to convergent, bifurcating, chaotic, or diverging behavior. Although often modeling a system with only very few nodal points, the IPF has already been shown to be highly precise. 

As the IPF already assumes a set of standpoints, which in the case of city planning are stakeholders, and can simulate very different types of temporal stakeholder developments as well as sudden phase changes, converging, oscillating, bifurcating, and chaotic behavior, it is used as a basis for a political and social prediction process. 

In the frappant case discussed at the beginning of the Introduction, six such stakeholders were identified. These are the artists, local stores, local residents, residents of Altona, the investor IKEA, and the City of Hamburg. Although the present paper outlines the basic framework of the model discussing real-world cases in detail is beyond its scope. Still,  the number of stakeholders used in the model is fixed to six throughout.

Such an iterative, nonlinear-dynamical process is scale-free, capable of modeling sudden phase changes, and includes convergent, oscillating, bifurcating, complex, and chaotic states. Scaling is a problem in social sciences in general, leading to so-called voltage drops when evidence found in one social group is scaled up to a broader level\cite{List2024}.

Like with the brain IPF, the social and political IPF suggested here needs adaptation of stakeholders. Adaptation is a notion with many aspects; for an overview of contemporary ideas, see\cite{Sawicki2023}. In the present case, adaptation is understood as a change of the internal state of a network participant or stakeholder according to its own state, and therefore, it is called self-adaptation.

The interaction between stakeholders in the model is formulated in two different versions: a fixed and an adaptive interaction. Therefore, the structural balance theory discussed above is not used as a model input but might emerge from it.

The IPF is an algorithm emerging from a Physical Culture Theory\cite{Bader2021}, assuming a common ground for humans, nature, culture, or technology by finding all to be electromagnetic fields when broken down to a physical basis. This includes consciousness, which is found to be a tempo-spatial electromagnetic field complex enough to arrive at conscious content of all kinds. Returning to this introductions beginning, where the self-organization nature of life in general was discussed, in this view, cultural artifacts like musical instruments, which are self-organizing structures by their very nature but made by humans, are viewed as enlarging the human body and mind with interacting artifacts as a way to maintain life by reducing a system entropy.

Therefore, as the IPF has already proven to work with musical instruments and the human brain, adding politics, social, and city planning issues to the model might help arrive at a model that bases culture and nature on a common ground.

\section{Method}

The social IPF assumes a set of interacting stakeholders. Each stakeholder s is represented by a system variable $g^s_t$ at discrete, adjacent time steps t. This system variable represents the relevant parameters of the stakeholder, which can be artistic freedom, health condition, financial situation, or the like. The IPF is always formulated from a standpoint where each standpoint has its own IPF. Here, each of $N$ stakeholders is a standpoint.

The IPF furthermore assumes a self-reference of each standpoint upon itself using a self-adaptation variable $\alpha^s_t$, adapting according to the standpoint itself. The damping variable $\alpha^s$ determines whether the system is stationary, oscillating, bifurcating, chaotic, and nonconverging.

In the follow, stationary time series are those with a stationary fixed point the system has converged to. Limit cycles lead to oscillatory behaviour. Bifurcations are present in cases of alternating values of the time series of different order. The lowest order bifurcations are those a time series alternates between two different, adjacent values, higher-order bifurcations are present when alternative values are more then two, typically four or eight. Finally, chaotic time series are highly complex, still not perfectly random.

This terminology has shown useful in musical applications, where a stationary time series corresponds to a musical sound of constant pitch, limit cycles are alternating pitches, so musical vibrato, while bifurcations are so-called multiphonics, where two are more musical pitches are present at the same time. Finally, chaotic behaviour is present during initial transients as well as in certain types of musical articulation like with low-pressure wind or bowed instrument sound production.

Such a system has extensively been studied with various forms of musical instrument acoustics, human-machine interaction, or brain dynamics. It is suitable for reconstructing the fundamental behavior of systems with only a few parameters to very high precision.

Furthermore, the interaction of the different stakeholders is modeled using an interaction variable $\beta^{s,i}_t$, of stakeholder i acting on the viewpoint stakeholder s at time point t. Again, $\beta^{s,i}_t$ is adapting at each time point.

In previous IPF formulation, e.g., with the classical guitar, each standpoint IPF could be calculated separately, as the values of $\alpha$ and $\beta$ were constant. In the case of the social IPF, where these constants are adaptive and, therefore, variables, the IPF can only be calculated simultaneously for all stakeholders or standpoints.

Social or political issues are often very complex and individual. To account for this, the proposed social IPF consists of three equations with three varying terms and a set of parameters.

\subsection{Social IPF system state equations}

The Impulse Pattern Formulation (IPF) was first motivated to describe sound propagation in musical instruments\cite{Bader2013}\cite{Linke2019a}. There, impulse-like shaped acoustic waves are emerging from strings or reeds which travel through the instrument, are reflected and damped, and return to their point of origin. In reed instruments a pressure impulse is creased at the mouthpiece, travels through the reed, is reflected at the bell, is damped through radiation and wall losses, and returns to the reed again to trigger a new impulse. With stringed instruments, the force of the string upon a soundboard is impulse-shaped too. It travels through plates, ribs or air volumes, is damped along the way only to return to the string point. It can be shown that only through this nonlinear dynamical behaviour of such systems a constant periodicity of the system appears which is the musical pitch.

Such a system can be described in its most simple form like

\begin{equation}
	e^{-\frac{\partial g}{\partial t}} = \frac{1}{\alpha} g \ ,
\end{equation}

with system parameter $g$, the impulse strength, and damping $\alpha$. The exponential models the damping of such a system, which is typically exponentially decaying in energy. In this formulation the precise shape of the impulse is not taken into consideration but only its strength is represented as system parameter g. Also, the IPF is taking a viewpoint, the point the impulse is sent out and returns to.

As impulses leave a viewpoint and return there at discrete time points. Therefore,  the system state $g_t$ is introduced like

\begin{equation}
	e^{-\frac{g_{t+1}-g_t}{\Delta t}} = \frac{1}{\alpha} g_t \ ,
\end{equation}

for adjacent times points t. Solving for $g_{t+1}$ leads to the iterative equation

\begin{equation}
	g_{t+1} = g_t - \log_e{\frac{g_t}{\alpha}} \ .
\end{equation}

Assuming systems with multiple reflection point, this equation can be written in a more general form which directly can be used as the social IPF like

\begin{equation}
	g^s_{t+1} = g^s_t - \log_e \left(\frac{1}{\alpha^s}\ g^s_t - \sum_{i=1}^{N, i \neq s} \beta^{s,i} e^{g^s_{t} - g^s_{t-d^{s,i}}}\right)\  + I^s_t.
	\label{Eq_basic}
\end{equation} 

Here $g^s_t$ is now defined for all stakeholders $s = 1,2,3,...N$ in the model. A time delay parameter $d^{s,i}$ with $i = 1,2,3,...N $ and $ i \neq s$ is necessary to define the time it takes until a stakeholder $i$ is acting on a stakeholder $s$. The stakeholder interaction strength $\beta^{s,i}$ again of stakeholder $i$ acting on stakeholder $s$ might vary according to the relation between stakeholders and therefore need to be a variable.

To account for unexpected external input to the system, like unpredicted ecological, political, or financial boundary condition changes, an external function $I^s_t$ is introduced, directly acting on the stakeholders. This implies adding or subtracting energy from the system, which is reasonable as no system exists in isolation.

In many cases, no such impact exists, and for one or all stakeholders s

\begin{equation}
	I^s_t = 0 \ .
\end{equation}

Therefore, $I^s_t$ is the first varying term that might or might not be included in a scenario.

\subsection{Stakeholder self-adaptation equation}

Adaptivity is needed due to the adaptive nature of living systems and has already been successfully implemented in the IPF model family in the brain IPF model\cite{}. Adaptivity of a self-referencing system can only be implemented using the system variable $g^s$ of the stakeholder itself and is suggested as a self-adaptation variable

\begin{equation}
	\alpha^s_{t+1} = \left(1/\alpha^s_t - C^s(g^s_t - g^s_{t-1})\right)^{-1} \ .
	\label{Eq_alpha}
\end{equation}

Here, $\alpha^s_t$ is time-dependent and altered according to the system's gradient $g^s$ with a constant $C^s$.

In some cases, a stakeholder might not adapt its self-reference, as in the case of laws or regulations acting in the system, which are not subject to any changes. In this case, for some stakeholder $s$, it is assumed that

\begin{equation}
	\alpha^s = const \ .
\end{equation} 

Therefore, $\alpha^s_t$ is the second varying term that needs to be decided when modeling a real social scenario.

\subsection{Stakeholder interaction equation}

\begin{figure}
	\centering
	\includegraphics[width=1\linewidth]{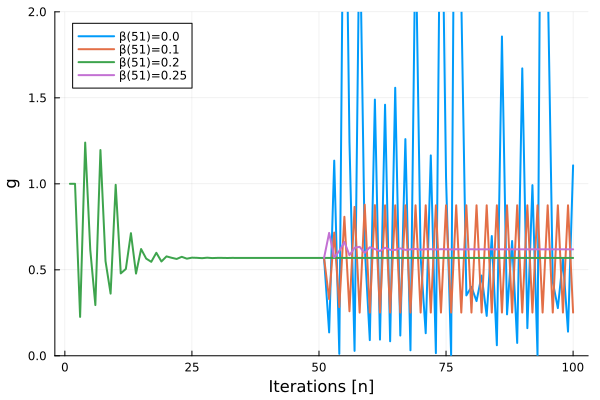}
	\caption{Time series of $g^s$ with $N=2$ interacting stakeholders and constant  $\alpha=0.369$ when suddenly changing $\beta^{s,i}$ at iteration step 51. $\beta^{s,i}$ always starts at $0.2$ and is suddenly increased to 0.25 (pink) converging to a new stationary limit or decreased to 0.1 or 0 leading to a low-order bifurcation or a chaotic behaviour respectively.}
	\label{fig:betaSudden}
\end{figure}

The impact of stakeholder $i$ on stakeholder $s$ is according to the interaction variable $\beta^{s,i}$. Changing $\beta^{s,i}$ during modeling is well known from the literature (e.g., \cite{Linke2019a,Linke2021a}). As shown in Fig. \ref{fig:betaSudden}, a sudden change of $\beta^{s,i}$ directly affects the system state $g$.

When keeping $\beta^{s,i} = 0.2$ (green) the stationary state is kept. Increasing $\beta^{s,i}$, after a short oscillatory transient phase $g^{s,i}$ converges to a new stationary limit (pink). Still, when decreasing $\beta^{s,i}$ to 0.1 a low-order bifurcation state is reached where $g1{s,i}$ alternates between two values (red). Decreasing $\beta{s,i}$ to zero leads to a chaotic behaviour (blue).

\begin{figure}
	\centering
	\includegraphics[width=1\linewidth]{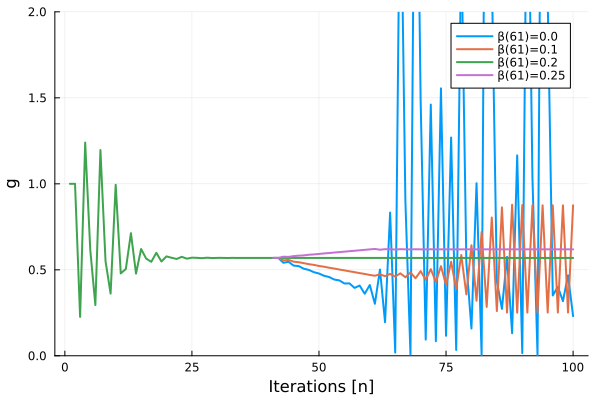}
	\caption{Time series of $g^s$ with $N=2$ interacting stakeholders and constant  $\alpha=0.369$ when gradually changing $\beta^{s,i}$ between iteration step 40 to 60. Start and end values of $\beta^{s,i}$ are the same as in Fig. \ref{fig:betaSudden} and the same system states are reached in the end. Still, during transition $g^{s,i}$ is stationary even if the fixed point is not.}
	\label{fig:betaCont}
\end{figure}

Nevertheless, as shown in Fig. \ref{fig:betaCont}, continuously changing $\beta^{s,i}$ keeps the system close to its fixed point. Thus, the system appears to be stationary in the sense of not displaying bifurcations or chaos even in regions where the fixed point is not stationary. However, as soon as $\beta^{s,i}$ does not change anymore, the system gradually shows the expected behavior.

The mutually changing of $\beta^{s,i}$ of different stakeholders can be subject to several interaction processes. In fixed interaction, the interaction variable  is proportional to the system variable of the impacting stakeholder $i$ multiplied by a constant $W^{x,i}$ like

\begin{equation}
	\beta^{s,i}_{t+1} = W^{s,i} g^i_{t-d^{s,i}} \ .
	\label{Eq_betaW}
\end{equation}

We refer to this version as fixed impact in the following\cite{Berner2023}\cite{Berner2021}.

Again, the time delay parameter $d^{s,i}$ of Eq. \ref{Eq_basic} is introduced to account for the delay one stakeholder acts on another.

Yet, another option is to assume that a stakeholder is actively altering the impact of another stakeholder upon him with respect to the difference between both system states, like

\begin{equation}
	\beta^{s,i}_{t+1} = \beta^{s,i}_{t} - V^{s,i} \left(g^s_{t} - g^i_{t-d^{s,i}}\right) \ ,
	\label{Eq_betaV}
\end{equation}

where $V^{s,i}$ is a constant analog to $W^{s,i}$.

This accounts for the fact that stakeholders are in constant interaction and try to balance their impact constantly in time from one to another. We refer to this version as adaptive impact in the following.

So $\beta^{s,i}_t$ is the third term to be decided for, therefore called Varying Term II below in a social scenario.

The time delay can be formulated alternatively like

\begin{equation}
	\beta^{s,i}_{t+1} = \beta^{s,i}_{t} - V^{s,i} \left(g^s_{t-d^{s,i}} - g^i_{t-d^{s,i}}\right) \ .
	\label{Eq_betaV_2}
\end{equation}

in its present state with a previous state of $g^i$ delayed by $d^{s,i}$ but in its previous state delayed by $d^{s,i}$

Here, the system state $g^s$ is not compared in its present state with a previous state of $g^i$ delayed by $d^{s,i}$ but in its previous state delayed by $d^{s,i}$. This would imply that one stakeholder remembers a previous state or that the impacting stakeholder i refers to a previous state of s. This is a more general version to be included, although it is not expected to be the general case. Still, we will discuss the impact of such a case below in some detail, too.

\subsection{Workflow of the social, political, and city planning IPF}
\label{sec:buildingIPF}

Therefore, the social IPF is a toolbox consisting of three equations, three varying terms, and three parameters. The three equations have been outlined above.

Tab. \ref{tab:varyingterms} gives an overview of the varying terms, again as discussed above. There might or might not be an external impact (Varying Term I) for each stakeholder $s$ separately. There might or might not be self-adaptation (Varying Term II) for each stakeholder separately. And interaction might be fixed or adaptive (Varying Term III). This last condition might not be exclusive, and both interactions might take place at the same time. In this paper, we first consider both terms alone to discuss their properties and leave a combined version for future studies.

The fixed parameters of the model are summarized in Tabl. \ref{tab:parameters}. The self-adaptation weights determine the self-adaptation strength of the variable $\alpha^s_t$, as are the fixed and adaptive interaction weights for the interaction variables $\beta^{s, i}_t$. The delays of stakeholders interacting with each other are fixed parameters.

\begin{widetext}
	\begin{center}
		\begin{table}
			\begin{tabular}{c|c|c|c}
				& Varying term I  & Varying terms II & Varying term III \\
				& \bf{External impact} & \bf{Self-adaptation} & \bf{Interaction} \\
				& $I^s_t$ &$\alpha^s_{t+1}$ & $\beta^{s,i}_{t+1}$ \\
				\hline
				Version I&External impact& Self-adaptation & Fixed interaction \\
				& variable &  $\alpha^s_{t+1}  \left(1/\alpha^s_t - C^s(g^s_t - g^s_{t-1})\right)^{-1}$ &  $\beta^{s,i}_{t+1} = W^{s,i} g^i_{t-d^{s,i}}$ \\
				\hline
				Version II&No external impact& No self-adaptation & Adaptive interaction \\
				& 0 & $\alpha^s_{t+1} = const$ & $\beta^{s,i}_{t+1} = \beta^{s,i}_{t} - V^{s,i} \left(g^s_{t} - g^i_{t-d^{s,i}}\right)$ \\
			\end{tabular}
			\caption{Varying terms of social IPF}
			\label{tab:varyingterms}
		\end{table}
	\end{center}
\end{widetext}

\begin{table}
	\begin{tabular}{c|c}
		Parameter & Symbol \\
		\hline
		Self-adaptation weight & $C^s$\\
		Fixed interaction weight & $W^{s,i}$\\
		Adaptive interaction weight & $V^{s,i}$\\
		Interaction delay & $d^{i,j}$
	\end{tabular}
	\caption{Parameters of social IPF}
	\label{tab:parameters}
\end{table}

In real-world situations, the social IPF is expected to predict a best practice for political action, giving an initial state of all variables and parameters and a desired time development of the system parameters for all stakeholders. To achieve this, it seems inappropriate to evaluate all possible combinations of varying terms, parameters, and initial variable states but pre-select varying terms and initial variables first to determine the best parameters to achieve the goal. In other words, political and social constraints determine the varying terms and parameters, while political and social actions are expressed in the parameters.

Modeling a social IPF takes the following steps:

\subsubsection{ Decision on stakeholders}

First, the stakeholders need to be determined. In the Frappant case, there would be six relevant parties. This also implies determining the system variable $g^s$ for each stakeholder. In the Frappant case, this might be artistic freedom and possibilities for the artists, well-being for the surrounding public, financial interest of the investor, or the like.

\begin{widetext}
	\begin{center}
		\begin{table}
			\begin{tabular}{|c|c|c|}
				\hline
				\textbf{Stakeholder< s}	& \textbf{Interest} & \textbf{Data/Measurements $g^s_t$} \\
				\hline
				Artists	& Artistic freedom &  qm rehearsal and performance spaces\\
				\hline
				Residents at Frappant	& Health condition &  Volume of traffic\\
				\hline
				Residents in neighbourhood	& Attractive shopping area &  Number of visitors (but/train users)\\
				\hline
				Local stores	& Increasing Rents/Take profits & Rent / financial balance \\
				\hline
				Administration/Government	& Social/financial balance & elections \\
				\hline
				Investors	& Take profits & financial balance \\
				\hline
			\end{tabular}
			\caption{List of stakeholders of Frappant example. Each stakeholder has an interest concerning the city planning project. This interest is transferred into quantitative figures like financial balance for investors or qm of performance or rehearsal space for artists over time as $g^s_t$ in the IPF.}
		\end{table}
		\label{tab:stakeholders}
	\end{center}
\end{widetext}

Tab. \ref{tab:stakeholders} lists the stakeholders, their interest in the city planning project and the data or measurements used as quantitative $g^s_t$ values in the model. So e.g. investors are interested in the financial balance, residents of Frappant are interested in their health condition which is threatened due to increased traffic, or artists fear they loose their performance and rehearsal spaces. 

All these figures are present as numbers and found in city data bases over the modeling time of the project and with a certain discretization time step. The highest resolution, e.g. traffic measured on a monthly base, is taken as smallest discretization time step.

Alternatively, the model starts with certain values and calculates these data over time into the future. 

\subsubsection{ Decision on varying terms}

For each stakeholder, a choice of the varying terms needs to be decided. In the case of Frappant, all stakeholders would be subject to self-adaptation; therefore, varying terms II would be a self-adaptation case.

\subsubsection{ Decision on initial values}

Each stakeholder's system variable $g^s$ has an initial value at the beginning of the process. This value needs to be decided in terms of the interpretation of the system variable $g^s$, might it be health state, artistic freedom, or social equality. 

Sometimes, the initial values' decision cannot be clearly determined perfectly. This is accounted for in step 5).

\subsubsection{ Decision on delay parameter}

Next, the delay parameter matrix $d^{s,i}$ of impacting all stakeholders onto all others must be set. This also implies a choice of the time step of the system for each iterative step. This time step will depend on the data available, like traffic measured in intervals of months, house rent costs known in steps of years, or traffic noise where data is available several times a day or even hours. 

The time step needs to be set to the smallest time interval necessary, which will determine the interaction's delays. Such interactions can be clearly set, like monthly payments, regular concert activity, fiscal seasons, or regular business meetings. Still, they can also be on a weekly or daily basis with buying decisions or depending on seasonal changes over the year.

Again, some delays might not be clearly determined, which is addressed in the next step.

\subsubsection{ Machine learning of possible scenarios}

After making the above decisions, the task is to find the parameters best suited to a desirable social or political output for all stakeholders. As the parameter space of the three parameters left to determine can become large, and as some of the above parameters might not be determined satisfactorily, a machine learning approach is taken.

For each parameter, and for all initial values of step 3) and delay parameters in step 4), reasonable upper and lower bounds are chosen. Also, for all varying parameters, a simulation step size that is small enough to reasonably discretize the parameter range from lower to upper bounds of each parameter is chosen. Such a parameter space consists of N simulations and can contain several thousand cases.

Each case is simulated, resulting in a time series over the desired total simulation time with a respective amount of time steps T for all stakeholders. Each stakeholder set of simulated system parameter $g^s_t$ values is taken as a feature vector of size $T \times s$.

A machine learning approach is used. In the present paper, we use a Kohonen self-organizing map\cite{Kohonen1995}\cite{Blass2019} for several reasons. First, after training and best-fit matching of the feature vectors to the map, the reason for such a fit can be found in the trained maps' feature vectors. Many machine learning algorithms, like deep-learning methods, do not easily allow to determine the reason for successful training. So, Kohonen maps are used here to understand the reason for possible cluster formation. Also, Kohonen maps do work with small data sets. Although these are not too often expected in the case of a social IPF, in some cases, this feature might be of benefit. Yet, a third reason is the visualization in a 2D map, which is beneficial again for estimating the amount and sizes of clustering.

\subsubsection{ Determining social and political strategies from social IPF prediction}

When a desired development of all stakeholders states $g^s$ is determined by political means, the trained machine learning algorithm is fed with the desired development. It gives best-fit parameters to fulfill the need. The social IPF, in this case, then predicts a particular social or political development and outcome for all stakeholders when the respective parameters are set.

Such a parameter setup also needs to be determined according to uncertain input spaces in initial values (step 3) or delay parameters (step 4). Therefore, stable clusters are required where, at best, all simulated cases within the uncertain parameter range meet the desired output.

\subsubsection{ Interpreting social or political actions from predicted parameters}

The parameters of self-adaptation weight $C^s$ and fixed or adaptive interaction weight $W^{s,i}$ or $V^{s,i}$ respectively, need to be translated into action or decisions.

With best-fit results in the case of adaptive interactions between several stakeholders demanding a high weight, stakeholders need to make sure both sides interact closely to achieve the appropriate goal. This might also determine a best-fit interaction delay, which could imply regular meetings, financial benefits, a suggested time interval for artistic or cultural activity, or the like.

\section{Results}

\subsection{Single stakeholder without self-adaptation with external disturbance}

To get familiar with the model, at first, the behavior of a single stakeholder, s = 1, with and without self-adaptation, with varying term II in version I and version II, is discussed, also including the disturbance of an external impact, so with a varying term I in version I. As an example, 50 iteration steps were performed with $\alpha = 0.56$, ensuring a converging system variable $g^1$ (first stakeholder). An initial $g_{1_0} = 1$ is set as boundary conditions.

\begin{figure}
	\centering
	\includegraphics[width=1\linewidth]{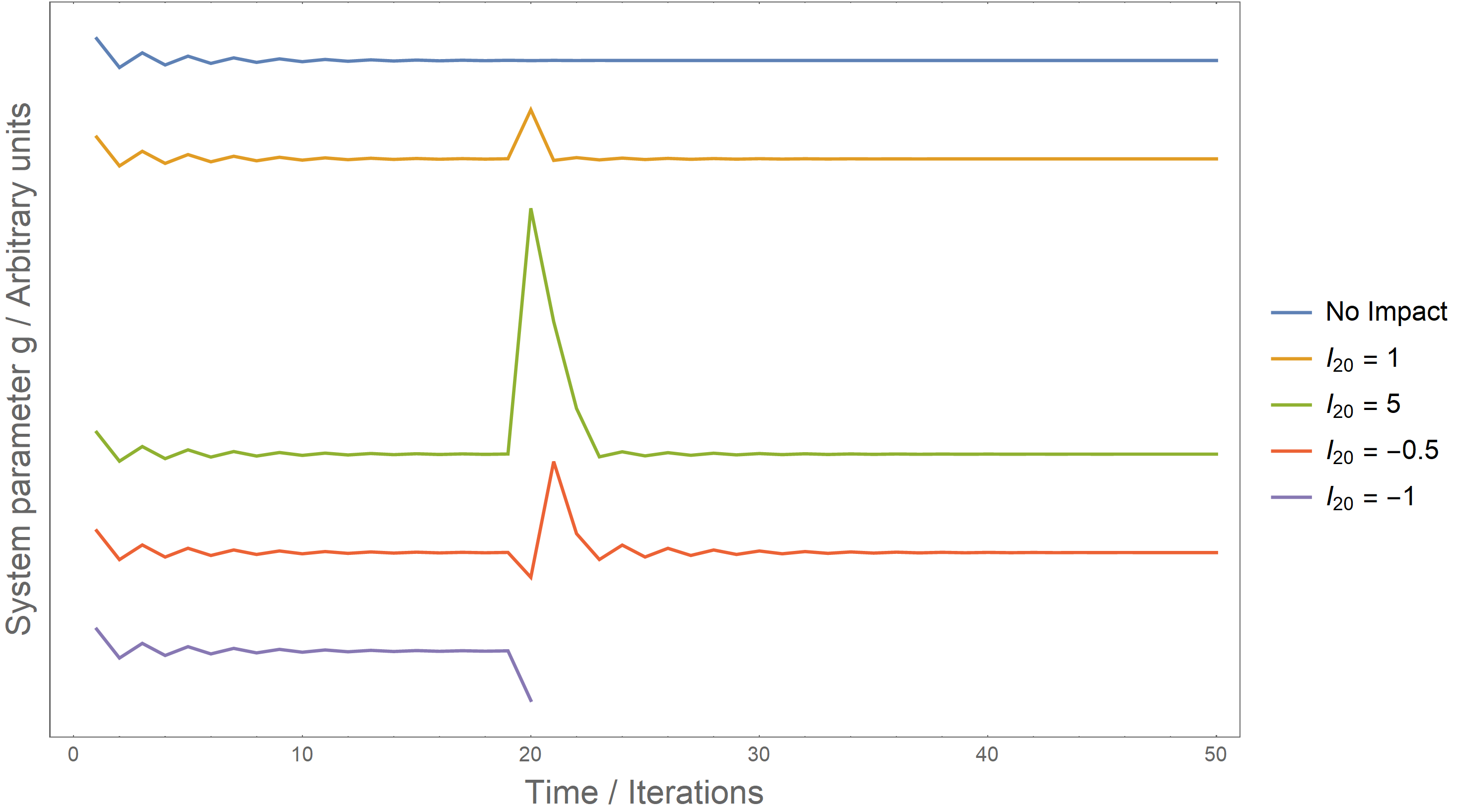}
	\caption{System variable $g^1$ of stakeholder one for $\alpha = 0.56$ and different impact $I_{20} = \{0, 1, 5, -0.5, -1\}$ at iteration point 20. Blue: no impact (varying term I version II), yellow positive impact $I_{20} = 1$  (varying term I version I) leading to a disturbance with returning convergence. Green: even stronger impact of $I_{20} = 5$ where convergence again appears only after a larger time interval. Red: negative impact $I_{20} = -0.5$ showing a strong positive reaction of the system followed by convergence. Purple: negative impact $I_{20} = -1$ leading to a negative $g^1$ and therefore a break down of the system.}
	\label{fig:ipfstadtplanungmodeli}
\end{figure}

Five cases are shown in Fig. \ref{fig:ipfstadtplanungmodeli}. The top blue curve shows $g^1$ over 50 iterations with $I_t = 0$ without impact. After an initial transient, $g^1$ becomes stationary and converges. 

The yellow curve shows the reaction to a positive $I_{20} = 1$, so an impact of one at a single iteration time 20. The system is disturbed and returns to its converging state. This is expected as such a distortion corresponds to a new initial $g_0$ after which convergence is ensured due to the choice of $\alpha = 0.56$ for which such a system is stationary.

The green curve shows a stronger impact of $I_{20} = 5$. Although the system becomes stationary again, it takes more time to recover.

The red curve shows an impact of $I_{20} = -0.5$, so a negative impact. Contrary to the positive impact, the system reacts with a strong positive $g^1$ at the next time step, from which it again converges.

Yet the purple curve shows an even larger negative impact of $I_{20} = -1$. Then $g^1$ becomes negative at the next time step, leading to a breakdown of the system.

All these behaviors represent reasonable reactions of a social system to an impact; it might converge to a stationary value, even with a negative impact, or it might break down and leave the stage.

Decreasing $\alpha$ does not change the system's reaction to an external impact considerably. Still, with low $\alpha$ around 0.37 for which chaotic behavior is reached, only a very small negative impact $I_t$ is needed to break the system down. A chaotic time series easily reaches low values of g where only a small negative impact makes g negative, leading to the breakdown.

In the Frappant case, the impact of the stakeholder City of Hamburg decided to tear down the building, representing a heavy impact on the artist's stakeholders. The City of Hamburg offered the artists a new building, the Victoria Kaserne, which allowed the artists to self-organize the building and its content. About 130 artists then founded a Fux e.G. and organized the artistic work with the low rental cost of the new building. The offering of the new building by the City of Hamburg only came into place due to strong political pressure implemented by the artists on the City of Hamburg to maintain artistic content and social life.

Therefore, the strong impact of one City of Hamburg stakeholder led to a strong movement of the artist's stakeholders in the City of Hamburg, which reacted to the artists again, and the whole system reached a new stability.

\subsection{Single stakeholder with self-adaptation with external impact}

\begin{figure}
	\centering
	\includegraphics[width=1\linewidth]{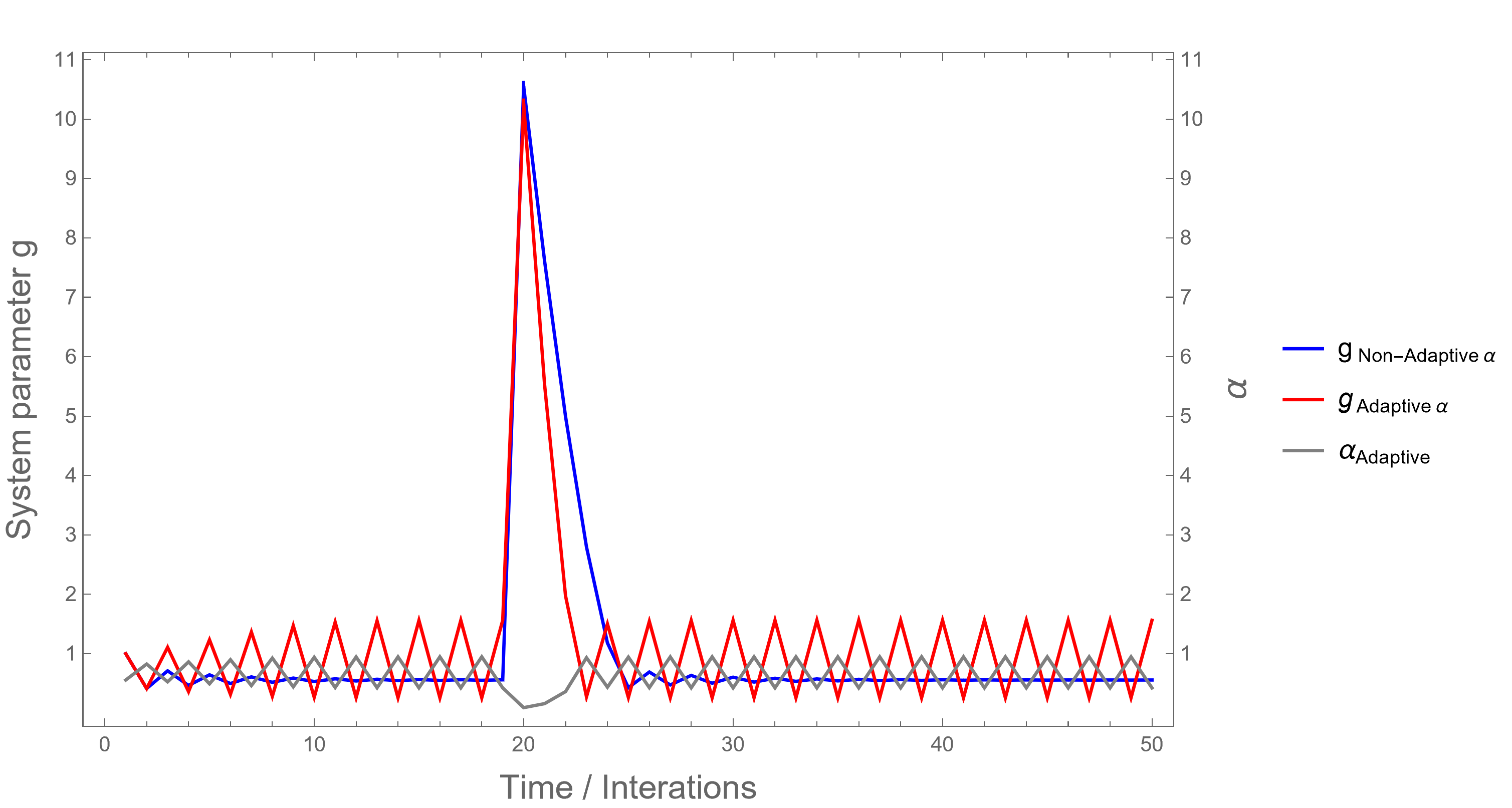}
	\caption{Adaptive version I of varying term II compared to non-adaptive version II concerning reaction of external input at iteration time point 20. Red: Version II time series from Fig. \ref{fig:ipfstadtplanungmodeli} with constant $\alpha = 0.56$. Blue: Version I time series with adaptive $\alpha$. Compared to Version II, the Version I time series is bifurcating due to adaptivity; still, it is returning to this state faster compared to Version II with static $\alpha$. Gray: Version I adaptive $\alpha$ corresponding to Version I system parameter $g^s_t$. After the impact on time point 20 $\alpha$ is way above the diverging limit of $\alpha \sim 0.37$, the system remains stable due to adaptation with a maximum $\alpha \sim 0.95$.}
	\label{fig:ipfstadtplanungmodeliibeispiel}
\end{figure}

To estimate the effect of self-adaptation, the model again with s=1 and a strong impact at iteration time 20 of $I_{20} = 10$ is used, but now including self-adaptation, so using varying term II in version I. Fig. \ref{fig:ipfstadtplanungmodeliibeispiel} compares $g^s_t$ for both cases, starting at $\alpha = 0.56$. In Fig. \ref{fig:ipfstadtplanungmodeli}, $\alpha$ = const, while now $\alpha$ is adaptive, therefore time-varying. Also now, $C^s = -1$ is used.

The red curve in Fig. \ref{fig:ipfstadtplanungmodeliibeispiel} is $g^s_t$ of varying term I version II already shown in Fig. \ref{fig:ipfstadtplanungmodeli}. The blue curve is the corresponding curve of varying term II version I, and the adaptive $\alpha$ is shown in gray. Both, $g^s_t$ and $\alpha_t$ are bifurcating but are stable in this respect. The impact at iteration time point 20 disturbs both curves considerably. Still, $g^s_t$ of version I returns to its bifurcating state much faster than $g^s_t$ of version II. 

So, as the adaptive version I of varying term II is much more active all through time, it is more feasible to return to a stable oscillation after a strong impact compared to the non-adaptive case. When reducing $C^s$, the bifurcating $g^s_t$ and $\alpha_t$ weaken their bifurcating amplitudes, and adaptation to a previous state is slower; the basic behavior remains the same.

To investigate this systematically, a parameter space is calculated for varying $C^s$ and $\alpha_0$ for three external impacts at iteration point 20 of $I^s_{20} = \{0, 1, 10\}$. The system runs over 100 iterations, and convergence is calculated, taking the mean of iterations 90 - 100 and only accepting real means < 100 and $\Im{g_t} = 0$ for all t. 

All calculations were performed using an initial system value $g^s_0 = 0.1$. This ensures $\Im{g_1} = 0$ for $\alpha_0 > 0.09048$ within the first iterative step, a value of $\alpha$ much lower than expected in the system at all. Using $g_0 = 1$ would lead to $\Im{g_1} \neq 0$ for $\alpha_0 \geq e$. Therefore, any adaptation algorithm would no longer be able to converge $\alpha_0 < 0.3679$, and extended stability by including any adaptation algorithm could not be tested. Using any $g_0$ is artificial in any case, as the initial time point of a social or political situation is arbitrarily set by an investigator while the stakeholders taking part in the social action most often existed before the chosen model starting time point.

\begin{figure}[!]
	\centering
	\includegraphics[width=.8\linewidth]{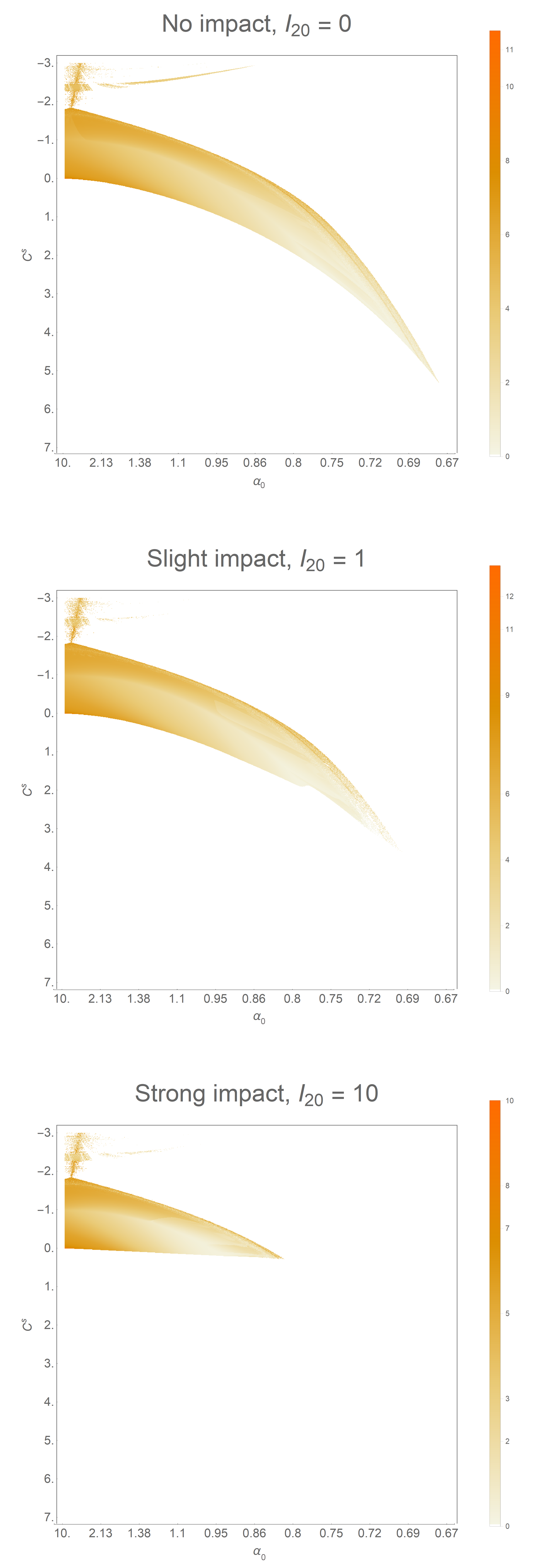}
	\caption{Convergence of adaptive Version I of varying term II for varying $C^s$ like $-3 \leq C^s \leq 7$ and initial $\alpha_0$ like 0.1961 $\leq \alpha_0 \leq$ 10 for three impact strength I. Top: $I_{20}$ = 0, no impact, $\alpha_0$ can be decreased way above the lowest value of non-adapting Version II with suitable $C^s$. Middle: reduced variability of $\alpha_0$; still, again, lower values are possible. Bottom: only negative $C^s$ lead to a stable solution and $\alpha_0$ only leads to conversion for values up to $\alpha_0 \sim 0.370$ as in the non-adaptive case.}
	\label{fig:ipfstadtplanungmodeliii0110}
\end{figure}

Fig \ref{fig:ipfstadtplanungmodeliii0110} shows the three cases for I = 0, so without adaptation, I = 1, and I = 10. The adaptation constant was varied from -3 $\leq C^s \leq 7$ and the initial $\alpha_0$ like 0.1961 $\leq \alpha_0 \leq$ 10.

In the top plot of Fig. \ref{fig:ipfstadtplanungmodeliii0110}, colors denote converging cases of $C^s$ vs. $\alpha_0$ combinations where the color corresponds to the mean converged value. A tail appears starting from $\alpha_0$ = 10 and 0 $\leq C^s \sim$ -2 which is getting narrower towards $\alpha_0 \sim 0.7$  and $C^s \sim$ 2. Therefore, choosing a suitable $C^s$ allows for much higher $\alpha_0$ to converge compared to the non-adaptive Version II with a minimum of $\alpha \sim 0.37$. Indeed, with decreasing $\alpha_0$, the range of $C^s$ for convergence is getting smaller, as expected. It is reliable that the convergence value is smaller towards higher $\alpha$ displayed by the lighter color as the IPF converges against $\alpha^s+\sum{\beta^{s,i}}$, which has already been shown analytically for the nonadaptive case\cite{Linke2019a}. 

Nevertheless, a cohesive area in which the IPF converges for all enclosed parameter values can always be found and is never fragmented in the sense of non-converging islands. Thus, convergence is ensured within the displayed regions.


Convergence is also present with negative $C^s$, as already discussed above. Still, negative $C^s$ allows only reduced $\alpha_0$ for convergence.

An unsystematic convergence area also exists for $C^s < -2$, which is quite speckled. Stakeholders in this region must be modeled very carefully. As the stable regions are narrow, small changes in the initial condition may drastically change the results. 

The middle plot of Fig. \ref{fig:ipfstadtplanungmodeliii0110} shows convergence for a slight impact $I_{20} = 1$. Again, lower $\alpha_0$ can be used to arrive at convergence compared to non-adaptive Version II., still slightly reduced compared to the case of $I_{20} = 0$.

Yet in the bottom plot with the case of $I_{20} = 10$, so strong impact, $\alpha_0$ can no longer be extended over the limit of Version II. This case also no longer allows for positive $C^s$.

Therefore, we can conclude that the adaptive Version I of varying term II is more stable over a wider range of $\alpha_0$ compared to Version II and, therefore, more flexible, as an expected consequence when including adaptation to the model.

Furthermore, with high external impact, the adaptation strength needs to be negative to have a robust, wide-range convergence. 

Therefore, the adaptive varying term II version I is superior to the non-adaptive version II as it covers all convergence of the non-adaptive with additional convergence capabilities with a suitable choice of parameters.

\subsection{Multiple stakeholders with fixed impact and no external disturbance}

Adding the interaction between multiple stakeholders leads to a temporal description of the system parameter $g^s$ for all stakeholders in the process. This is implemented by the interaction parameter $\beta^{s,i}_t$ of stakeholders i impacting on stakeholder s. First, fixed impact is discussed, which means including the Version I of Varying term III as shown in Tab. \ref{tab:varyingterms}.

When modeling city planning, a starting time point needs to be defined, and the initial values for all stakeholders $g^s$ and the parameters need to be decided. Different scenarios can happen. There might be a running system, like with the Frappant case, where many stakeholders have already reached convergence, and only one or two new stakeholders come in, or a new impact disturbs the system. In present-day city planning, this might be the more likely case, with upcycling or densifying of cities being the main focus. However, a system might also be set up from scratch, like when planning a new area of construction. 

In this simulation, we assume all stakeholders are already present in the system before the simulation starts at $t=0$. Still, couplings $\beta^{s,i}$ are often not known or hard to determine, which leads to the simplification of uncoupled stakeholders before $t=0$. Thus Eq. \ref{Eq_basic} and \ref{Eq_betaW} are valid $\forall \ t>=0$. $\forall \ t<0$ the system can be described by the IPF with $\beta^{s,i}=0 \forall \ s, i$. As we assume the system to be running for a sufficiently long time, all stakeholders have already converged to their fixed point. When assuming that no $\beta$ are involved before $t=0$ the system is at the fixed point $g^s_f=\alpha^s$ according to Eq. \ref{Eq_basic} and thus $g^s_{t}=\alpha^s \ \forall t<0$. Then, all necessary initial values of the simulation are set. Further, the time delay $d^{s,i}=i$ directly depends on the couped stakeholder.

In the following, IPFs with $N=6$ stakeholders are used throughout, as taken from the Frappant example. Each IPF is iterated for 500 steps using Equations \ref{Eq_basic} and \ref{Eq_betaW}. $d^{s,i} =1$ is used throughout. By randomly varying $\alpha^s$, IPFs were calculated until 1000  non-diverging time series were obtained. This amount was reached after 3523 attempts. Thereby, $\alpha^s$ was chosen within a range of $\alpha^s \in [0.2, 1]$. Also, the coupling constant $W^s_i=W$ was fixed for all stakeholders but varied for each IPF again randomly in the range of $W \in [0.01, 1/6]$.

According to the process of predicting a best practice of political action according to stakeholders' needs in a public participation process, possible scenarios need to be identified, as discussed above. Therefore, a clustering algorithm is needed to determine typical temporal developments of stakeholders $g^s$. If this is successfully implemented, the inverse problem is solved, starting from a desired output and deriving parameters of the system leading to this output. From these parameters, political actions are derived as part of the workflow discussed above in Section \ref{sec:buildingIPF} of city planning or political action.

The clustering algorithm used is a self-organizing Kohonen map (SOM) for the reasons discussed above. The SOM is trained with the 1000 converging IPFs. Thereby, the time series of each stakeholder $s$ is normalized to mean=0 and std=1. Thus, the SOM organizes the data according to the overall shape of $g_t$ but not on the absolute values, as they are highly dependent on $\alpha$. We, therefore, arrive at types of $g^s$ developments.

\begin{figure}[!]
	\centering
	\includegraphics[width=1\linewidth]{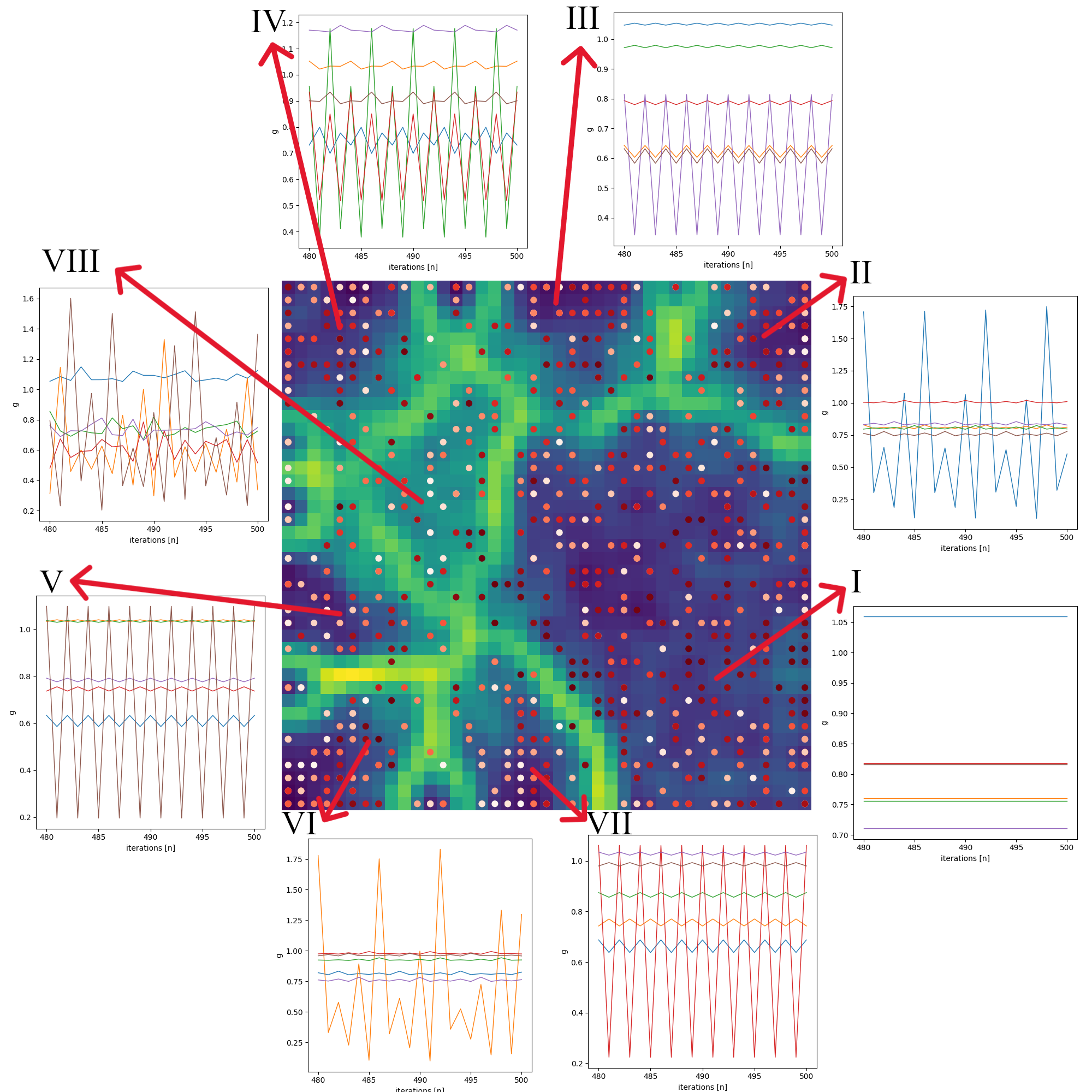}
	\caption{Trained SOM of 1000 IPF cases of six stakeholders using fixed interaction with Version I of Varying terms II. The background color is the u-matrix displaying similarities (dark colors) or differences (light colors) of neighbouring neurons. Therefore, the yellow ridges are boundaries between clusters present in darker regions. Each dot is an IPF $g^s$ iteration', which is placed according to its best fit to the feature vectors of the trained SOM. The colors of the dots refer to the coupling strength $W \in [0.01, 1/6]$; lighter/darker colors correspond to weaker/stronger coupling. For each of the clusters or regions I to VIII, an assigned time series $g^s_t$ is picked and plotted aside the SOM as a typical example of stakeholders $g^s$ development of the last 20 values of the 500 iteration steps for the respective regions I to VIII.}
	\label{fig:SomW}
\end{figure}

Fig. \ref{fig:SomW} shows the trained SOM for Version I of Varying term III. The background color is the so-called u-matrix, displaying similarities and differences of neighboring neurons. Therefore, the Euclidean distances of a neuron to its neighbors are calculated in the feature space. Then, the mean value of these distances allows an evaluation of how far away the neurons are located in the feature space and thus, how different they are.

In Fig. \ref{fig:SomW}, lighter yellow colors have strong differences, and darker blue colors have small differences. Therefore, the yellow ridges are boundaries between blue regions of very similar neurons. The dots represent the 1000 IPF cases the map was trained with, and they are positioned after training at the best-fit neurons for each IPF. The colors of the dots show varying coupling strengths $W$, where lighter colors are weaker and darker colors have stronger coupling.

\begin{figure}[!]
	\centering
	\includegraphics[width=1\linewidth]{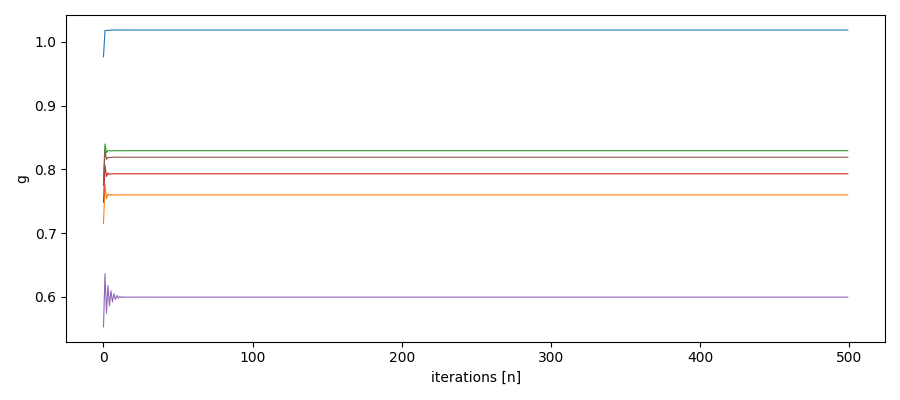}
	\caption{Example IPF time series of region I of Fig. \ref{fig:SomW} of $g^s$ converging for all N=6 stakeholders, calculated from Eq. \ref{Eq_betaW}. The initial adaptation vector was $\alpha^s \approx \{0.98, 0.71, 0.79, 0.75, 0.55, 0.77 \}$ and $W\approx 0.011$. Therefore, the largest region I of all eight regions is that of convergence of all stakeholders.}
	\label{fig:timeSomW_1}
\end{figure}

In this SOM, clearly, eight distinct regions of the u-matrix are visible and named I to VIII, with I being the largest region. In region I, all stakeholders converge to their fixed point. A typical $g^s$ time series of all six stakeholders is shown in Figure \ref{fig:timeSomW_1}. After an initial short transient phase, all $g^s$ converge. Region I is the only region where such convergence takes place.

Such a situation could be after construction of a new housing area. Although at the planning stage negotiations will happen, leading to the complex initial state in the simulation, it is expected that for some time after construction ends, the inhabitants and all other stakeholders are satisfied.

\begin{figure}[!]
	\centering
	\includegraphics[width=1\linewidth]{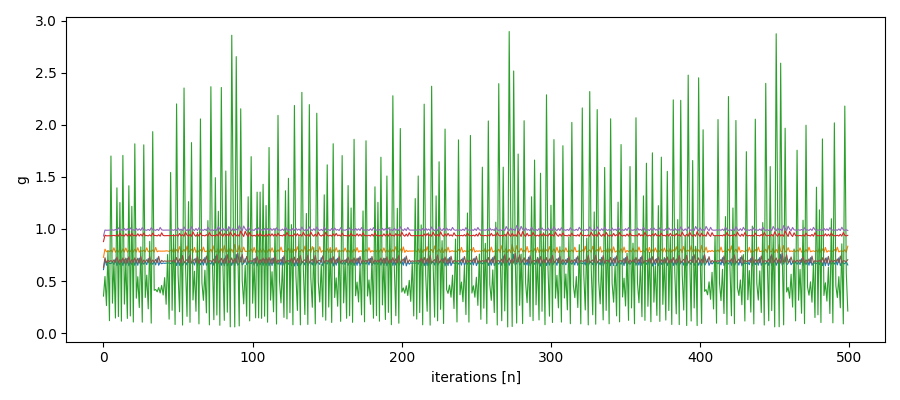}
	\caption{Example IPF time series of region VIII of Fig. \ref{fig:SomW} of a chaotic time series of stakeholders $g^s$, as calculated from Eq. \ref{Eq_betaW} with initial stakeholder $\alpha^s \approx \{0.61, 0.73, 0.36, 0.88, 0.93, 0.63 \}$ and $W\approx 0.016$.} 
	\label{fig:timeSomW_8}
\end{figure}

The region \textit{VIII} in Fig. \ref{fig:SomW} is also unique. Here, the stakeholders $g_s$ show chaotic time series and complex patterns. An example IPF $g^s$ time series is shown in Fig. \ref{fig:timeSomW_8}.

Typical recent examples are big projects within cities where construction already began and due to financial or political struggle progress only very slowly or not at all. This leads to uncertainties for investors, entrepreneurs, local residents, politics, city planners, etc. The IPF could help here to determine in such cases possible solutions. 

\begin{figure}[!]
	\centering
	\includegraphics[width=1\linewidth]{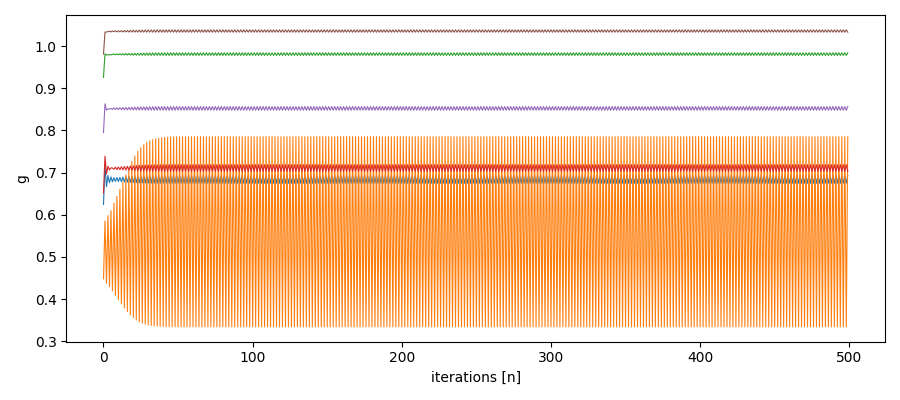}
	\caption{Example IPF time series of region VI as a bifurcating time series of $g^s$ as calculated by Eq. \ref{Eq_betaW} with initial stakeholders $\alpha^s \approx \{0.62, 0.44, 0.93, 0.65, 0.79, 0.98 \}$ and $W\approx 0.014$.}
	\label{fig:timeSomW_4}
\end{figure}

The remaining regions \textit{II} to \textit{VII} are rather similar. All lead to bifurcations. However, they differ in the order of bifurcation as well as in the magnitude of fluctuation of the different stakeholders. Another difference between those regions is that different stakeholders dominate in terms of their stability or their bifurcations. An example of such a time series is shown in Fig. \ref{fig:timeSomW_4}. There, five stakeholders converge while one stakeholder is bifurcating.

Typical such cases are ecology left in a very uncertain or devastated state, although the other stakeholders are satisfied. As discussed above, this is one of the main challenges of our times.

\begin{figure}[!]
	\centering
	\includegraphics[width=1\linewidth]{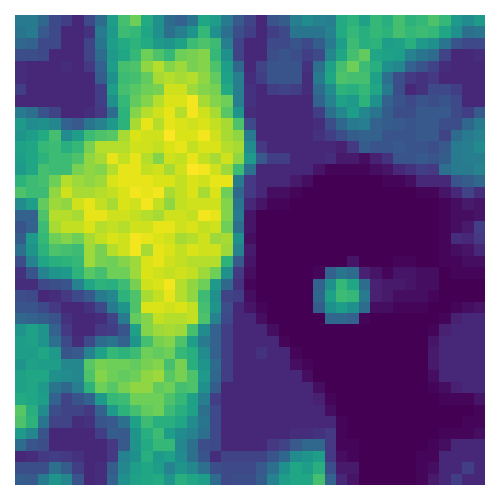}
	\caption{Number of different values $g^s_t$ when inspecting the last ten values of each stakeholder time series. Shown for every neuron of the trained SOM of Fig. \ref{fig:SomW}. Values are assumed to be different when they differ more than 5~\% from each other. Dark/light regions refer to low/high numbers of different values.
		Region I clearly has the lowest values. As it is converging throughout, no fluctuations are detected. Also, the chaotic region VIII has the highest values. Still, region II - VII can now be differentiated with respect to the number of different values and, therefore, the order of bifurcation.}
	\label{fig:SomW_component}
\end{figure}

A more systematic view of regions \textbf{II} - \textbf{VII} can be obtained from Fig. \ref{fig:SomW_component}. Here, for each node of the SOM, the last ten values of the time series $g^s_t$ are inspected in terms of the number of different values. Values are assumed to be different when they differ more than 5~\% from each other. One can clearly distinguish the stable region \textit{I} and the chaotic region \textit{VIII}. Furthermore, \textit{II} - \textit{VII}, while looking rather similar in Figure \ref{fig:SomW}, are found to differ in terms of the order of bifurcation.

Using only version I of varying terms III of fixed interaction between stakeholders shows all kinds of possible scenarios for city planning. These include

\begin{itemize}
	\item Stationary, Convergence: All stakeholders arrive at their desired value.
	\item Bifurcation: Complex output of very different kinds appear, like convergence of many stakeholders, while one or two stakeholders show complex development. All stakeholders bifurcate to a certain extent with all kinds of bifurcating orders.
	\item Chaotic: Stakeholders are highly chaotic, and system output is unpredictable.
	\item Non-converging: City planning has failed as the system breaks down and stakeholders disappear.
\end{itemize}

So, the inverse problem of finding parameters and therefore, political actions to arrive at a desired output for all stakeholders is solved using the proposed method. Still, a second interaction process will appear between stakeholders, that of adaptive interactions discussed below.

\subsection{Multiple stakeholders with adaptive impact and no external disturbance}

Along the method of fixed interaction discussed above, for adaptive interaction again IPFs for $N=6$ stakeholders with 500 iteration steps each were calculated, using Eq. \ref{Eq_basic} and now Version II of Varying terms III of Eq. \ref{Eq_betaV}.

This time, we also take an even closer look at the initial conditions. Also, the delay term $d^{s,i}$ of stakeholder i acting on stakeholder s as amount of iterations is inspected more closely. 

In general, four different cases are possible:

\begin{itemize}
	\item \textbf{I1:}
	Same as the previous modeling example of fixed impact, it is assumed that no coupling is present $\forall t<0$ and that the system has already settled itself, and thus $g^s_{t}=\alpha^s \ \forall t<0$. At $t=0$, suddenly, all stakeholders are coupled to each other.
	
	\item \textbf{I2:}
	Similar to \textbf{I1}, but as the system is supposed to be uncoupled $\forall t<0$, it is not realistic that all stakeholders are suddenly coupled to each other. The IPF models the impact of the coupled stakeholders by the damped impact of preceding system states $g^s_{t-d^{s,i}}$. As coupling is only present for $t\geq0$, previous system states can not act on future ones. Thus, the different $\beta^{s,i}_t$ are only taken into account if $t \geq d^{s,i}$.
	
	\item \textbf{I3:}
	It is assumed that the system is not settled at the start. Thus $g^s_{t}$ may have an arbitrary value $\forall t<0$. Here, for simplicity, we chose $g^s=1$. Like in \textbf{I1} it is assumed that all stakeholders are suddenly coupled to each other at $t=0$.
	
	\item \textbf{I4:}
	Again, like in \textbf{I3} $g^s_{t}=1 \ \forall t<0$. However, similar to \textbf{I2} the different $\beta^{s,i}_t$ are only taken into account if $t \geq d^{s,i}$.
	
\end{itemize}

For each of the above initial conditions, 125 different, non-diverging time series are calculated, where again $\alpha^s \in [0.5, 1]$ are chosen randomly. The coupling $V^s_i=V$ is supposed to be the same for all stakeholders. Values are also chosen randomly in the range of $V \in ]0, \ 0.01]$. The time delays are different for each stakeholder 

\begin{equation}
	d^{s,i}= i C_{dt}^s \, 
\end{equation}

where the scaling factor is the same for all stakeholders and chosen randomly $\{ C_{dt}^s=C_{dt} \in \mathbb{N} : 0<d \leq 16 \}$.

The resulting time series are shown in Fig. \ref{fig:timeSomV_i1} to \ref{fig:timeSomV_i4} as examples for each of the four initial conditions. They are very much different from those of the fixed interactions discussed above. Bifurcations and chaos are rarely present, rather damped oscillations appear. Also, due to the mutual adaptation of $\beta^{s,i}_t$, all coupled IPFs converge to the same fixed point. This fixed point might be reached in an exponential development, like shown in Fig.\ref{fig:timeSomV_i3}, it might show vibrations before convergence, like in Fig. \ref{fig:timeSomV_i1} - \ref{fig:timeSomV_i2}, or it might show very slow convergence, so basically each stakeholder keeps its own state.

An example is a legally binding land-use plan process (B-Plan Verfahren) in which all stakeholders negotiate over time on a planning of a area. The result is planning certainty. Due to negotiations, an adaptive and self-reference process is leading to perfect convergence of all stakeholders needs. Of course, this process is not the realization, which again can lead to complex situations like financial shortages or the like.

To analyze the models' behavior systematically and for an inverse problem solution, again, a self-organizing Kohonen map (SOM) is trained. This time, no distinguished regions or clusters can be detected, as was the case with fixed interaction when training the map with the simulated time series of $g^s$. This is expected when examining Fig. \ref{fig:timeSomV_i1} - \ref{fig:timeSomV_i4}. The differences in the time series are gradual. Still, different phenotypes appear, like exponential convergence, vibrations, or slow convergence, with practically different values for each stakeholder.

Therefore, to examine these types, a few meaningful parameters must be derived from the modeled time series $g^s$. At first, each series is divided into three windows of equal length to distinguish between its start, its end, and its middle part. For each time window, three values are calculated:

\begin{itemize}
	\item \textbf{Convergence detection:}
	
	The overall standard derivation of the system states of all stakeholders in a certain time window $g^s_t \forall s \in [0,6],t \in [t_n,t_m]$ is calculated to detect systems convergence and if differences between stakeholders $g^s_{t_n}$ is still present. 
	
	\item \textbf{Oscillation detection:}
	
	The ratio between the actual $g$ and the theoretical fixed point \cite{Linke2019a} $g_f=\alpha+\sum \beta_k$  is calculated for each stakeholder individually. When the standard deviation of this ratio is calculated, a first impression of whether damped oscillations occur can be derived. The SOM is then trained by the mean of this ratio over all stakeholders.
	
	\item \textbf{Transient evaluation:}
	
	The absolute value of Pearson Correlation $r=\frac{conv(X,Y)}{\sigma_X \sigma_Y}$ is calculated. With $conv(X,Y)$ being the covariance of the two Vectros $X,Y$ and $\sigma_X, \sigma_Y$ being the standard derivation of the two Vectors. 
	
	If $X$ is the time series of system states of a stakeholder $g^s_t$ and $Y$ a straight line $f(t)=t$, $r^s$ describes whether the $g^s_t$ is monotonous falling or growing, or if wave-like fluctuations occur, as shows up in Fig. \ref{fig:timeSomV_i1}. By calculating the mean value of the absolute value of $r^s$ over all stakeholders the transient behavior can be evaluated.
	
\end{itemize}

For each of the 4 $\times$ 125 time series, these three values were calculated and fed as feature vectors into a SOM. 

\begin{figure}[!]
	\centering
	\includegraphics[width=1\linewidth]{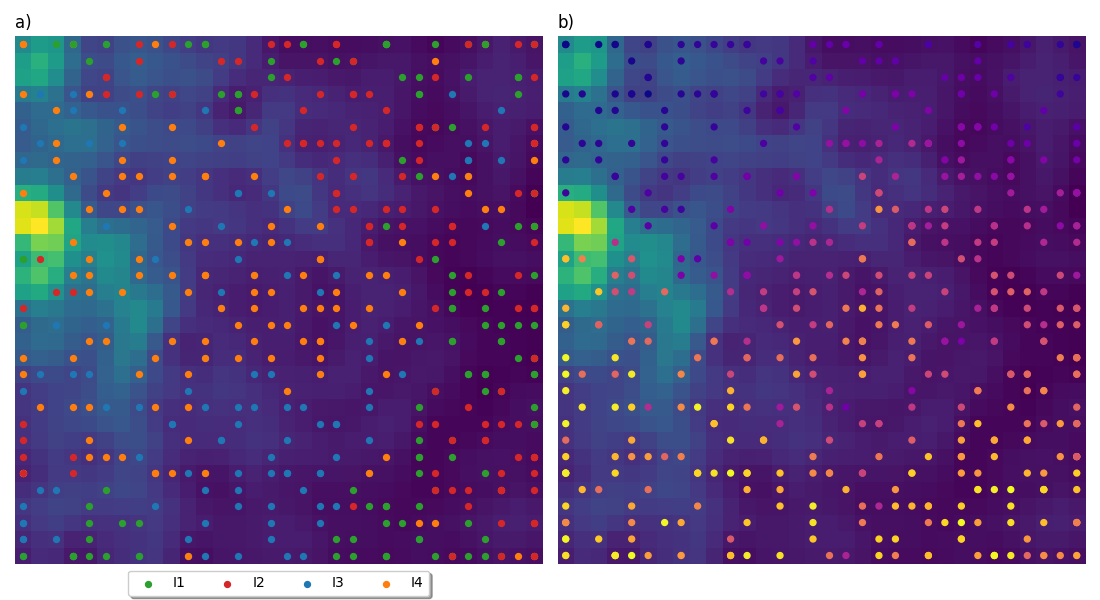}
	\caption{Trained SOM for adaptive interaction between stakeholders using Version II of Varying term III of Eq. \ref{Eq_betaV}: a) and b) show the same u-matrix, but the dots, representing the sorted training data are colored following different criteria: In a), the colors are chosen due to the initial conditions. While green dots refer to \textbf{I1}, red dots refer to \textbf{I2}, blue dots refer to \textbf{I3}, and orange dots refer to \textbf{I4}. In b), the colors represent the coupling strength $V$, while darker values refer to weaker coupling.}
	\label{fig:SomVred}
\end{figure}

\begin{figure}[!]
	\centering
	\includegraphics[width=1\linewidth]{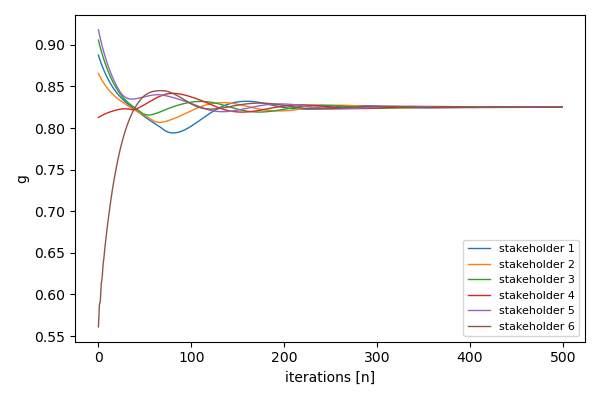}
	\caption{Example time series of $g^s$, for each of the N=6 stakeholders, resulting from an IPF model with initial condition \textbf{I1} and $\alpha^s \approx \{0.89, 0.87, 0.91, 0.81, 0.92, 0.56 \}$, $V\approx 0.01$ and $C_{dt}=12$. On the SOM shown in Figure \ref{fig:SomVred}, this time series is sorted into the lower left corner.}
	\label{fig:timeSomV_i1}
\end{figure}

\begin{figure}[!]
	\centering
	\includegraphics[width=1\linewidth]{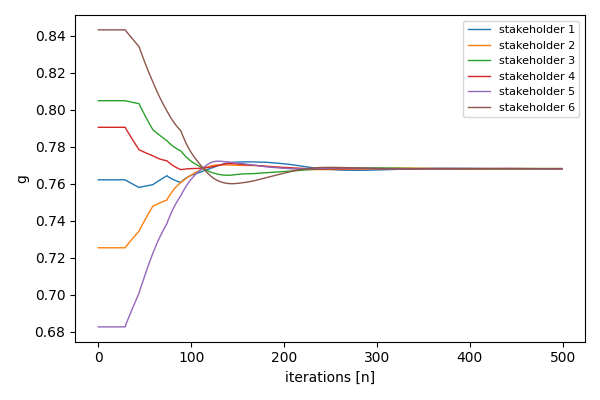}
	\caption{Example time series of $g^s$, for each of the N=6 stakeholders, resulting from an IPF model with initial condition \textbf{I2} and $\alpha^s \approx \{0.76, 0.73, 0.80, 0.79, 0.68, 0.84 \}$, $V\approx 0.008$ and $C_{dt}=15$. On the SOM shown in Figure \ref{fig:SomVred}, this time series is sorted into the lower right corner.}
	\label{fig:timeSomV_i2}
\end{figure}

\begin{figure}[!]
	\centering
	\includegraphics[width=1\linewidth]{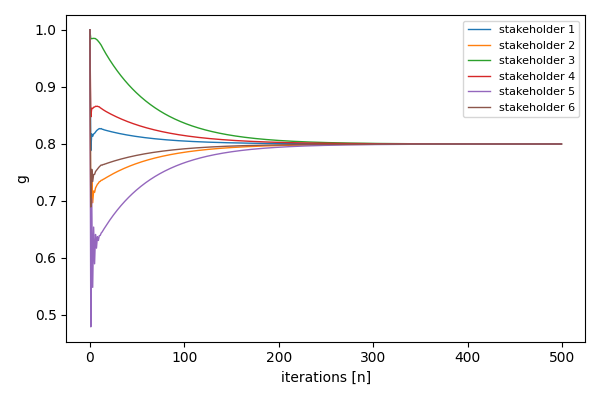}
	\caption{Example time series of $g^s$, for each of the N=6 stakeholders, resulting from an IPF model with initial condition \textbf{I3} and $\alpha^s \approx \{0.89, 0.87, 0.90, 0.81, 0.92, 0.56 \}$, $V\approx0.003$ and $C_{dt}=2$. On the SOM shown in Figure \ref{fig:SomVred}, this time series is sorted into the small island of orange and blue dots, surrounded by red and green dots, in the upper right of the map.}
	\label{fig:timeSomV_i3}
\end{figure}

\begin{figure}[!]
	\centering
	\includegraphics[width=1\linewidth]{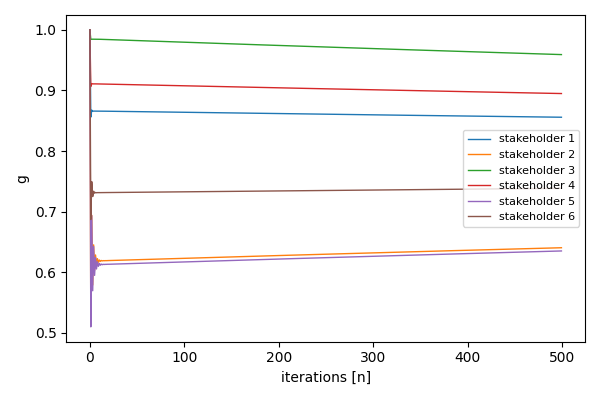}
	\caption{Example time series of $g^s$, for each of the N=6 stakeholders, resulting from an IPF model with initial condition \textbf{I4} and $\alpha^s \approx \{0.86, 0.62, 0.98, 0.91, 0.61, 0.73 \}$, $V\approx4.7\times 10^-5$ and $C_{dt}=2$. On the SOM shown in Figure \ref{fig:SomVred}, this time series is sorted into the upper left corner.}
	\label{fig:timeSomV_i4}
\end{figure}

Fig \ref{fig:SomVred} shows this trained SOM using the values as input feature vectors to the neural net. Again, the background color is the u-matrix, which now does not show clear regions or clusters. This points to the continuous nature of the $g^s$ in all conditions. 

Fig. \ref{fig:SomVred} has the same u-matrix in plots a) and b). Also, the $g^s$ of all cases are best-fit to respective neurons in the same way. The difference between both plots is in the coloring of the dots. While a) colors are according to initial conditions I1 - I4, in b) the colors are chosen according to the coupling strength $V$ of the adaptive terms of $\beta^{s,i}$ in Eq. \ref{Eq_betaV}.

Differentiating with respect to initial conditions \textbf{I1} - \textbf{I4} in Fig. \ref{fig:SomVred} a) shows systematic differences between initial conditions. Only arbitrary $g_0$ of  \textbf{I3} and  \textbf{I4} show chaotic transients, with an examples shown in Fig. \ref{fig:timeSomV_i3} and Fig. \ref{fig:timeSomV_i4}). As soon as the $g$ is close to $g_f$, the system will behave initially stationary.  As Eq. \ref{Eq_betaV} changes $\beta$ iteratively, fluctuations are prevented, and the system always shows smooth behavior as already shown in Fig. \ref{fig:betaCont}.

\begin{figure}[!]
	\centering
	\includegraphics[width=1\linewidth]{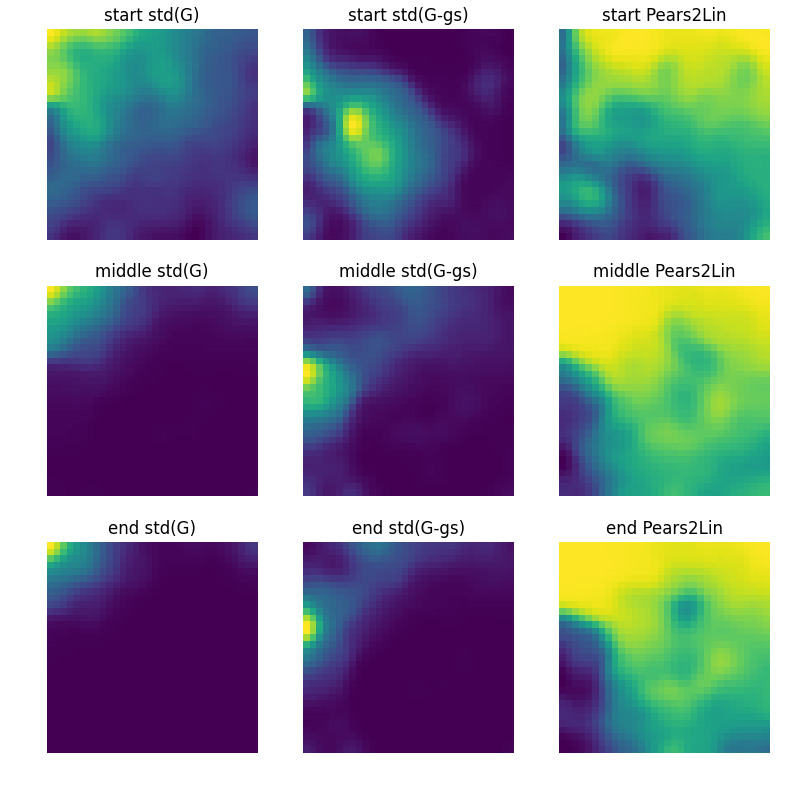}
	\caption{Component planes of trained SOM of Fig. \ref{fig:SomVred} of the adaptive interaction case, Version II of Varying term III when using Eq. \ref{Eq_betaV}. The rows refer to the different time windows. The columns denote the three detection methods of convergence, oscillation, and transient evaluation as input feature vectors to the SOM and are displayed as magnitudes of these different features.}
	\label{fig:SomVred_component}
\end{figure}

This phenomenon is also strongly influenced by the coupling strength $V$. Fig. \ref{fig:SomVred} b) shows that the SOM also sorts training data according to $V$. The reason for this becomes clear when examining the component planes of the trained SOM, shown in Fig. \ref{fig:SomVred_component}. Columns show the distribution of one feature vector entry and one component over the trained SOM of the three analysis parameters: convergence and oscillation detectors as well as the transient evaluation. By correlating the component planes to the dot distribution, the reason for sorting the $g^s$ on the SOM can be found.

In the SOM in Fig. \ref{fig:SomVred} b), the dots in the upper part of the plot are darker, pointing to a weak coupling. The transient evaluation component planes on the right of Fig. \ref{fig:SomVred_component} show bright values in their upper parts, corresponding to large values of the transient evaluation and thus linear behavior. Fig. \ref{fig:timeSomV_i4} shows that, due to weak coupling, changes of $\beta$ are small, such that only minor changes in $g$ appear. The system is very stationary but takes a long time to converge. When the coupling is strong, like in Fig. \ref{fig:timeSomV_i1} or \ref{fig:timeSomV_i2}, the $g^s$ approach each other very fast, but might overshoot, leading to vibration behavior. This is detected by the transient evaluation method.

The length of the vibration also depends on the use of delay Fig. \ref{fig:timeSomV_i2}: When the delay $d^{s,i}$ is large, it takes several iteration steps until the vibration starts, and usually even more for it to end. Further, a large $d^{s,i}$ allows a more complex and chaotic time series.

\subsubsection{Complex delay interactions}

In the methods Section, we presented a more complex version of the adaptive interaction. Eq. \ref{Eq_betaV_2} describes the change of $\beta^{s,i}$ as caused by a previous state of stakeholder $i$ on a previous state of stakeholder $s$, not on the present state of stakeholder $s$ like in Eq. \ref{Eq_betaV}. This adds a memory to the stakeholders as they remember their previous states and take this into account while judging the impact of a coupled stakeholder.

\begin{figure}[!]
	\centering
	\includegraphics[width=1\linewidth]{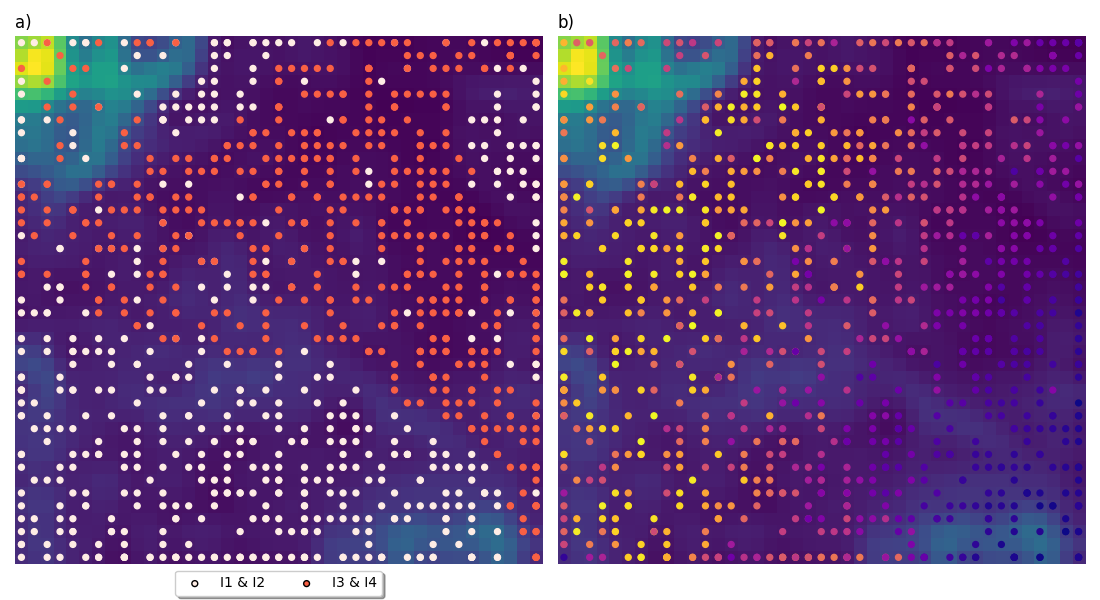}
	\caption{Trained SOM of adaptive interaction but with more complex delay term using Eq. \ref{Eq_betaV_2}: a) and b) show the same u-matrix, dots represent the best-fit IPF cases onto the map. a) Colors are chosen according to initial conditions. Red dots: \textbf{I1} and \textbf{I2}, white dots: \textbf{I3} and \textbf{I4}. b) Colors represent coupling strength $V$, with darker colors representing weaker, and lighter colors representing stronger coupling.}
	\label{fig:SomV}
\end{figure}

Calculating again 125 time series for each of the four initial conditions \textbf{I1} - \textbf{I4} and using them to train a SOM leads to results shown in Fig. \ref{fig:SomV}.

Again, the different models are ordered according to the coupling strength $V$, as visible in Fig. \ref{fig:SomV} b). Furthermore, Fig. \ref{fig:SomV} a) shows that a general difference between initial conditions \textbf{I1} \& \textbf{I2} and \textbf{I3} \& \textbf{I4} remains. Therefore, the complex delay does not change the system's behavior fundamentally.

\begin{figure}[!]
	\centering
	\includegraphics[width=1\linewidth]{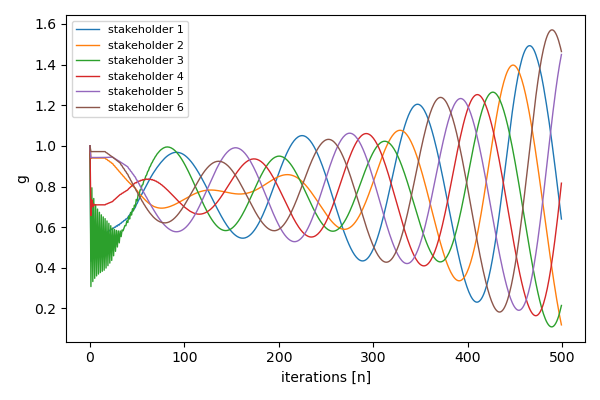}
	\caption{Time series of $g^s$, for each of the N=6 stakeholders of the adaptive interaction Version II of Varying terms III but here with complex delay of Eq. \ref{Eq_betaV_2}  with initial condition \textbf{I3}, $\alpha^s \approx \{0.57, 0.94, 0.50, 0.71, 0.94, 0.97 \}$, $V\approx0.008$, and $C_{dt}=8$. The case is mapped to the upper left corner of the SOM shown in Figure \ref{fig:SomV}.}
	\label{fig:timeSomV_chaos}
\end{figure}

Still, the system is more unstable. Fig. \ref{fig:timeSomV_chaos} shows that using Eq. \ref{Eq_betaV_2}, the chaotic transients can be significantly extended, and the IPF may not converge anymore.

\section{Conclusions}

The proposed algorithm is able to model all typical kinds of city planning, social, or political time developments of interacting and self-adapting stakeholders. These include conversion to a common fixpoint, conversion to different states, bifurcations or oscillations of many kinds, as well as chaotic behaviour. The reaction to an external impact is also realistic with adaptation, convergence, or breakdown.

The workflow of city planning includes the solution of the inverse problem, which is finding a set of parameters needed to arrive at a desired output. The model suggests different such scenarios, estimates their stability in terms of changing external conditions, and points to problems for certain stakeholders within the project. 

Using a self-organizing Kohonen map to arrive at such suggestions enables modelers to find the reasons for successful parameter estimations. Also, the whole field of possible parameters and interactions is described in the two-dimensional map representation, giving an overview of potential alternatives to a proposed output or parameter space.

The model is flexible in terms of using different varying terms in different versions. This allows flexible modification of the model according to real-world situations. Also, by adding or omitting terms, certain kinds of interactions or self-adaptations might be suggested by the system, therefore making aware of critical stakeholder interaction processes needing more care during a planning process.

The proposed city planning model is only exemplified in this paper using typical city planning situations. As it has now proven to be powerful, robust, and flexible for planning processes, the next step is to validate the model with respect to a complex real-world situation. Such a case could be a finished project where the stakeholder developments are known. It might also be a running project where the IPF is performed in parallel. This is beyond the scope of the present paper, which first needs to outline the basic behavior of such a system. Still, this IPF has been developed in cooperation with city planners and is, therefore, close to meeting their constraints and demands. Future work will be conducted, giving examples of real-world models.

Then, a vision for the future would be to implement an IPF in all big planning processes, leading to much more efficient and robust plannings that take care of all stakeholder's demands and needs.

\section*{Data Availability Statement}

Data available on request from the authors.

\begin{acknowledgements}
	This research was funded under the Program "Innovative Hochschule" (innovative university) by the Federal Ministry of Education and Research (BMBF) of Germany (Grant No.13IHS232C) and the City of Hamburg.
\end{acknowledgements}

\bibliography{aipsamp}

\end{document}